\documentclass[preprint,aps,showpacs,prl,
floatfix,12pt,amsmath,amssymb]{revtex4}
\usepackage{graphicx}
\usepackage{dcolumn}
\usepackage{bm}
\usepackage{epsfig}
\usepackage{color}

\begin{document}

\title{Complete spin polarization of degenerate electrons in semiconductors
near ferromagnetic contacts}
\author{A.~G.~Petukhov$^{1}$, V.~N.~Smelyanskiy$^{2}$, and
V.~V.~Osipov$^{2,3}$}
\date{\today }
\affiliation{$^{1}$Physics Department, South Dakota School of Mines and
Technology, Rapid
City, SD 57701 \\
$^{2}$NASA Ames Research Center, Mail Stop 269-3, Moffett Field, CA 94035 \\
$^{3}$Mission Critical Technologies, Inc., El Segundo, CA 90245 }

\begin{abstract}
We show that spin polarization of electron density in nonmagnetic degenerate
semiconductors can achieve 100\%. This effect is realized in
ferromagnet-semiconductor $FM-n^{+}$-$n$ junctions even at moderate spin
selectivity of the $FM-n^{+}$~contact when the electrons are extracted from
the heavily doped $n^{+}-$semiconductor into the ferromagnet. We derived a
general equation relating spin polarization of the current to that of the
electron density in
nonmagnetic semiconductors. We found that the effect of the complete spin
polarization is achieved near $n^{+}$-$n$ interface when an effective
diffusion coefficient goes to zero in this region while the diffusion
current remains finite.
\end{abstract}

\maketitle

Combining carrier spin as a new degree of freedom with the established
bandgap engineering of modern devices offers exciting opportunities for new
functionality and performance. This emerging field of semiconductor physics
is referred to as semiconductor spintronics \cite{zutic2004a,awschalom2002a}%
. The injection of spin-polarized electrons into nonmagnetic semiconductors
(NS) is of particular interest because of the relatively large
spin-coherence lifetime, $\tau _{s}$, and the promise for applications in
both ultrafast low-power electronic devices \cite%
{zutic2004a,awschalom2002a,datta1990a,sato2001a,jiang2003a,osipov2004a,bratkovsky2004a}
and in quantum information processing (QIP) \cite%
{awschalom2002a,nielsen2000a,taylor2002a,taylor2003a,xiao2004a}. Main characteristics
of the spin injection are the spin polarizations of the electron density $%
P=\left( n_{\uparrow }-n_{\downarrow }\right) /n=\Delta n/n$ and the current
$\gamma =\left( j_{\uparrow }-j_{\downarrow }\right) /j=\Delta j/j$. The
value of $\gamma $ determines a magnetoresistance ratio and performance of
spin-valve devices \cite%
{datta1990a,rashba2000a,fert2001a,bratkovsky2004a,osipov2004a}. The value of
$P$ determines polarization of the recombination radiation measured in most
of the experiments of optical detection of spin injection \cite%
{hanbicki2002a,hanbicki2003a,ohno2003a}. Moreover, a high value of $P$ is
crucial for QIP devices \cite%
{awschalom2002a,taylor2002a,taylor2003a}. It has been implied in most of the
previous theoretical works on spin injection \cite%
{aronov1976a,johnson1987a,johnson2003a,rashba2000a,fert2001a,yu2002a,yu2002b,albrecht2002a,bratkovsky2004b,osipov2004c}
that $P$ cannot exceed $\gamma $. This assumption complies with existing
observations in which different magnetic materials such as magnetic
semiconductors or ferromagnetic metals (FM) have been used as injectors of
spins into semiconductors \cite{zutic2004a,awschalom2002a}.

It follows from a formal consideration by Yu and Flatte \cite%
{yu2002a,yu2002b} that $P$ can, in principle, exceed $\gamma $ in
nondegenerate semiconductors when electron spins are extracted from NS into
FM (reverse bias). However, more detailed studies by Osipov and Bratkovsky
\cite{bratkovsky2004a,osipov2004a}, taking into account tunneling through a
Schottky barrier in simple FM-NS junctions, revealed that $P<\gamma $ due to
a feedback formed during the tunneling process. The condition $P<$ $\gamma $
holds for both nondegenerate and degenerate semiconductors and for both
reverse- and forward- biased simple FM-NS
junctions \cite{osipov2004c,bratkovsky2004b,
osipov2005a}.

\begin{figure}[tbph]
\includegraphics[width=.75\linewidth,angle=270]{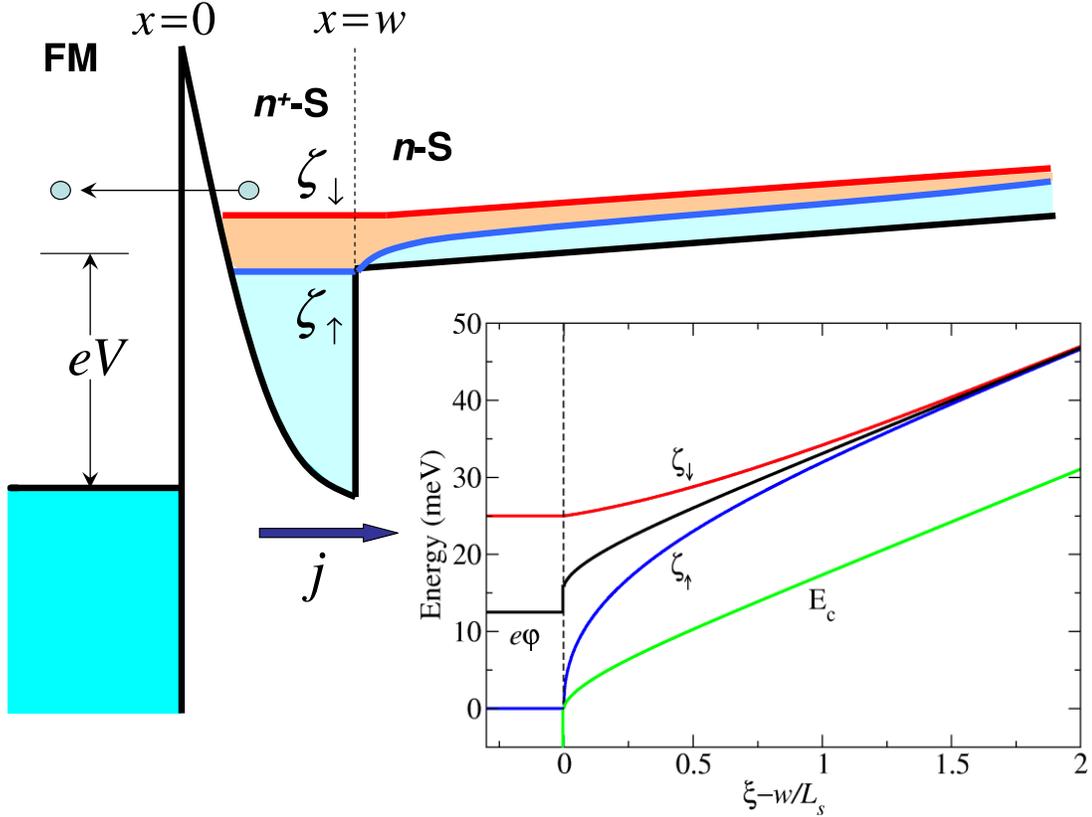}
\caption{(Color online.) Schematic view of the proposed FM-$n^{+}-n$
heterostructure. 
Inset: calculated
$\protect\zeta_\protect\sigma$
and electrostatic potential $\protect\varphi$ at $j=j_{c}$}
\label{structure}
\end{figure}

Unlike previous works where simple FM-NS junctions were studied, in this
Letter we consider a band-engineered FM-$n^{+}$-$n$ structure containing a
thin super-heavily doped $n^{+}$-layer and a degenerate semiconductor $n$%
-region (Fig.~\ref{structure}). The effect in question is based on spin
extraction and nonlinear dependence of the nonequilibrium spin density on
the electric field. A non-equilibrium electron gas becomes completely spin
polarized when a quasi-Fermi level for one type of
carrier (e.g.  $\zeta _{\uparrow}$ ) reaches the bottom of the conduction
band $E_{c}$ near $n^{+}$-$n$ interface. The spin extraction from NS was
predicted by I.~Zutic~\textit{et al.}~ \cite{zutic2002a} for forward-biased $%
p-n$ junctions containing a magnetic semiconductor, was studied in detail
for FM-NS junctions \cite{bratkovsky2004a,osipov2004a}, and was
experimentally found in forward-biased MnAs/GaAs Schottky junction \cite%
{stephens2003a}. However, both the predicted and observed values of the spin
polarization $P$ were rather small.

Let us consider a nonmagnetic semiconductor in which non-equilibrium
spin-polarized electrons are described by the quasi-Fermi distribution and
the current density $j_{\sigma }$ can be expressed as:
\begin{equation}
j_{\sigma }=e\mu n_{\sigma }E+eD_{\sigma }\frac{\partial n_{\sigma }}{%
\partial x}=\mu n_{\sigma }\frac{\partial \zeta _{\sigma }}{\partial {x}},
\label{j}
\end{equation}%
where $E$ is the electric field, $\mu =e\tau _{p}/m$ is the mobility, $\tau
_{p}$ is the momentum relaxation time, $m$ is the effective mass and $%
D_{\sigma }$ is the diffusion coefficient of electrons with spin $\sigma $.
Here we used the generalized Einstein's relation \cite{ashcroft1976a}
$\mu =eD_{\sigma }
{\partial \ln (n_{\sigma })}/{\partial \zeta _{\sigma }}.$
 Since $\mu $ is spin independent this relation shows that $D_{\sigma }$ does
not depend on spin in nondegenerate semiconductors ($n\propto \exp (-\zeta
_{\sigma }/kT)$) while the spin dependence of $D_{\sigma }$\ is crucial for
degenerate semiconductors.

Using Eq.~(\ref{j}) and $j=const$ in steady-state we find:
\begin{equation}
\gamma =P+(1-P^{2})\frac{\mu n}{2j}\frac{\partial \Delta \zeta }{\partial x}
\label{gamma3}
\end{equation}%
where $\Delta \zeta =\zeta _{\uparrow }-\zeta _{\downarrow }$.
It follows that $\gamma =P
$ only in the absence of diffusion. Therefore the term $P$\ in this equation
can be interpreted as a spin-drift term while the other one as a
spin-diffusion term.

\begin{figure}[tbph]
\vskip -10 pt \includegraphics[width=.8\linewidth]{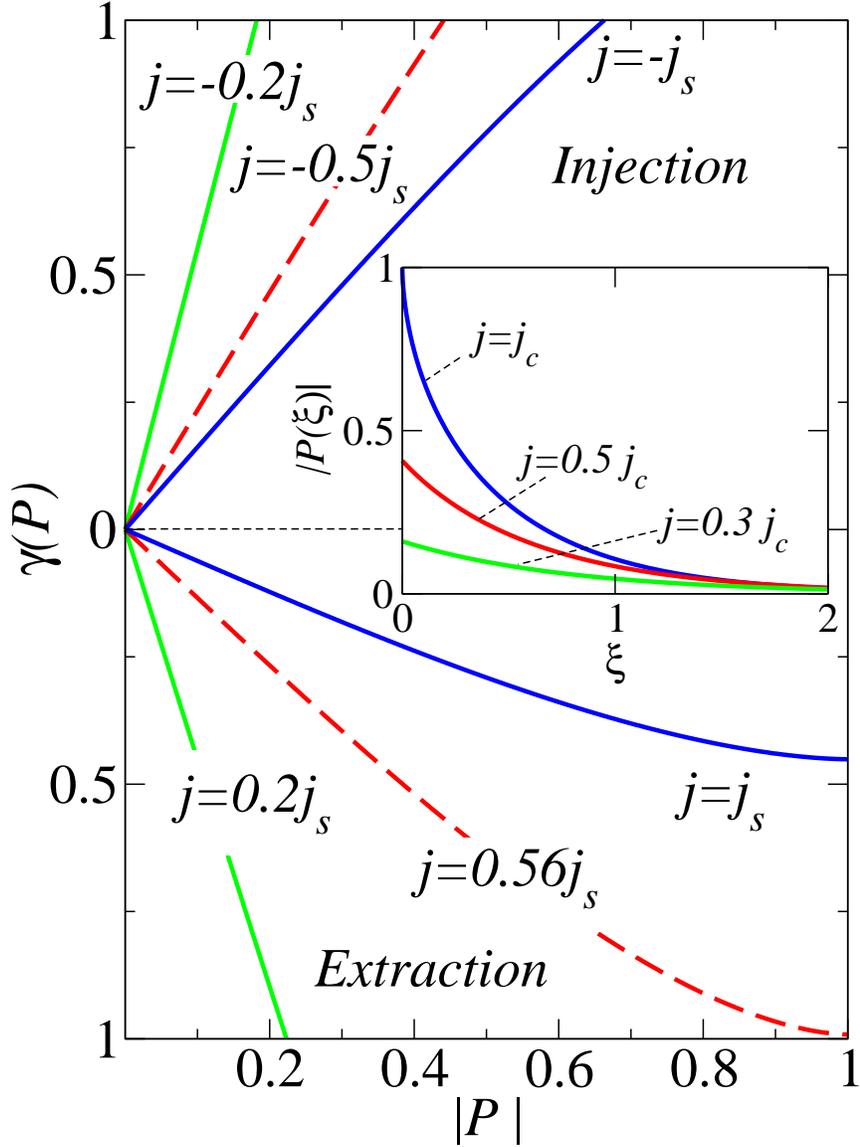} \vskip -10 pt
\caption{(Color online.) Solutions of Eq.~({\protect\ref{gammaP}}) for different $j$. Inset: spatial distribution of $P(\protect\xi )$
}
\label{gamma_P}
\end{figure}

We consider the FM-$n^{+}$-$n$ structure shown in Fig.~\ref{structure}. We
note that the sign of $\gamma >0$ is independent of the direction of the
current while the sign of $P$ does depend on it. Namely, $P>0$ for
reverse-biased junctions (total current $j=\left( j_{\uparrow
}+j_{\downarrow }\right) <0$) when the spin injection occurs and $P<0$ for
forward-biased junctions ( $j>0$) when the spin extraction takes place. Due
to very high \ electron density we use the electro-neutrality condition, $%
n=n_{0}$, where $n_{0}$ is the equilibrium electron density. In this case
Eq.~(\ref{gamma3}) reduces to:
\begin{equation}
\gamma =P+\frac{en_{0}D(P)}{j}\frac{dP}{dx},  \label{gamma4}
\end{equation}%
where
\begin{equation}
D(P)=\frac{\mu }{2e}(1-P^{2})\frac{d\Delta \zeta (n_{0},P)}{dP}  \label{DP}
\end{equation}%
is a bi-spin diffusion coefficient. For degenerate semiconductors at low
temperatures
\begin{equation}
\Delta \zeta =E_F\left[ (1+P)^{2/3}-(1-P)^{2/3}\right] ,
\label{zi-}
\end{equation}%
where $E_F=mv_F^2/2$ is  the equilibrium Fermi
energy and 
$v_{F}=(\hbar /m)(3\pi ^{2}n_{0})^{1/3}$. 
From Eqs.~(\ref{DP})-(\ref{zi-}) we obtain \cite{osipov2005a}
\begin{equation}
D(P)=(v_{F}^{2}\tau _{p}/{3})\tilde{D}(P),  \label{DPdeg}
\end{equation}%
where
\begin{equation}
\tilde{D}(P)=\frac{1}{2}(1-P^{2})^{2/3}\left[ (1+P)^{1/3}+(1-P)^{1/3}\right]
\label{dimsd}
\end{equation}%
The steady-state continuity equation for spin-dependent currents in NS
reads:
\begin{equation}
\frac{dj_{\sigma }}{dx}=\frac{e}{2\tau _{s}}\left( n_{\sigma }-n_{-\sigma
}\right) .  \label{continuity}
\end{equation}%
Introducing the spin diffusion length $L_{s}^{2}=(v_{F}^{2}\tau _{p}\tau
_{s})/3$ and dimensionless length $\xi =x/L_{s}$, from Eqs.~(\ref{continuity}%
) and (\ref{gamma4}) we find:
\begin{equation}
\frac{j}{j_{s}}\frac{d\gamma }{d\xi }=P,\text{ where }\gamma =P+\frac{j_{s}%
\tilde{D}(P)}{j}\frac{dP}{d\xi }  \label{dimcon}
\end{equation}%
and $j_{s}=en_{0}L_{s}/\tau _{s}$. We remember that $j$ is positive for spin
extraction and negative for spin injection. System of Eqs.~(\ref{dimcon})
leads to a non-linear drift-diffusion equation in dimensionless variables:
\begin{equation}
\frac{d}{d\xi }\left( \tilde{D}(P)\frac{dP}{d\xi }\right) +\frac{j}{j_{s}}%
\frac{dP}{d\xi }-P=0,  \label{dims}
\end{equation}%
Eliminating $dP/d\xi $ from Eq.~(\ref{dimcon}) we can reduce the
second-order non-linear equation (\ref{dims}) to the first-order equation
relating $\gamma $ to $P$:
\begin{equation}
\frac{j^{2}}{j_{s}^{2}}\frac{d\gamma }{dP}=\tilde{D}(P)\frac{P}{\gamma -P}
\label{gammaP}
\end{equation}%
We notice that $\tilde{D}(P)=const$ in nondegenerate semiconductors and
solution of Eq.~(\ref{gammaP}) reduces to the known result \cite%
{yu2002a,yu2002b,osipov2004a}. A boundary condition for Eq.~(\ref{gammaP})
can be obtained from the asymptotics $P\rightarrow 0$ and $dP/d\xi
\rightarrow -P/l_{j}$ at $\xi \rightarrow \infty $, where $l_{j}^{-1}=\sqrt{%
j^{2}/4j_{s}^{2}+1}+j/2j_{s}$. Using these asymptotics and Eqs.~(\ref{dimcon}%
) we find the boundary condition for Eq.~(\ref{gammaP}):
\begin{equation}
\lim_{P\rightarrow 0}{\gamma }/{P}=1-j_{s}/jl_{j}  \label{boundary}
\end{equation}%
Solution of Eq.~(\ref{gammaP}) with this boundary condition is a \emph{%
universal} function $\gamma (P,j)$ determining local relation between $%
\gamma $ and $P$. Numerical solutions of Eq.~(\ref{gammaP}) in the domain $%
0\leq |P|\leq 1$ are shown in Fig.~\ref{gamma_P} for different values of $j$%
.

The parameter $l_{j}$ is a dimensionless spin penetration length \cite%
{aronov1976a,yu2002a,yu2002b,osipov2004a}. This length at large currents
tends to infinity for spin injection ($j<0$) or to zero for spin extraction (%
$j>0$). It means that the spin accumulation layer is expanded away from the
interface under spin injection and compressed towards the interface under
spin extraction. As it follows from Eq.~(\ref{boundary}) $\gamma
/P\rightarrow 1$ at $j\rightarrow -\infty $ and $\gamma /P\rightarrow 0$ at $%
j\rightarrow \infty $. A solution with $|P|=1$ does not exist for $j<0$ and $%
\gamma <1$ but is possible for positive $j\geq 0.56j_{s}$ (see Fig.~%
\ref{gamma_P}). Therefore, the spin extraction in forward-biased FM-S
junctions provides a possibility to create a 100\% spin-polarized,
non-equilibrium electron gas in a non-magnetic semiconductor near the FM-S
interface. In this Letter we demonstrate that this possibility is feasible
and technologically sound. Solutions of Eq.~(\ref{dims}) for spatial
distributions of $P(\xi )$ in $n-S$-region (Fig.~\ref{structure})
are shown in the inset to Fig.~\ref{gamma_P}. The
function $|P(\xi )|$ reaches 1 at the interface  and becomes
singular when $j=j_c>0.56j_{s}$.
The value of $j_c$ depends on boundary conditions and will be calculated
below.
Our numerical analysis shows that at this point $|P|=1-C(j)(\xi-w/L_s)
^{3/5}$, where $C(j)=1.145+0.549j/j_{s}$. The current spin polarization at $%
\xi =w/L_s$ equals to:
\begin{equation}
\gamma =(3/5)(j_{s}/j)C(j)^{5/3}-1  \label{gamma7}
\end{equation}%
It follows from Eq.~(\ref{gamma7}) that $|P|$ reaches 1 when $\gamma <1$
provided that $j>0.56j_{s}$. One can see from
Fig.~\ref{gamma_P} that the value of $|P|=1$ can be achieved at rather
small $\gamma $ if the current is sufficiently large.

\begin{figure}[tbph]
\label{Pj} \includegraphics[width=.8\linewidth]{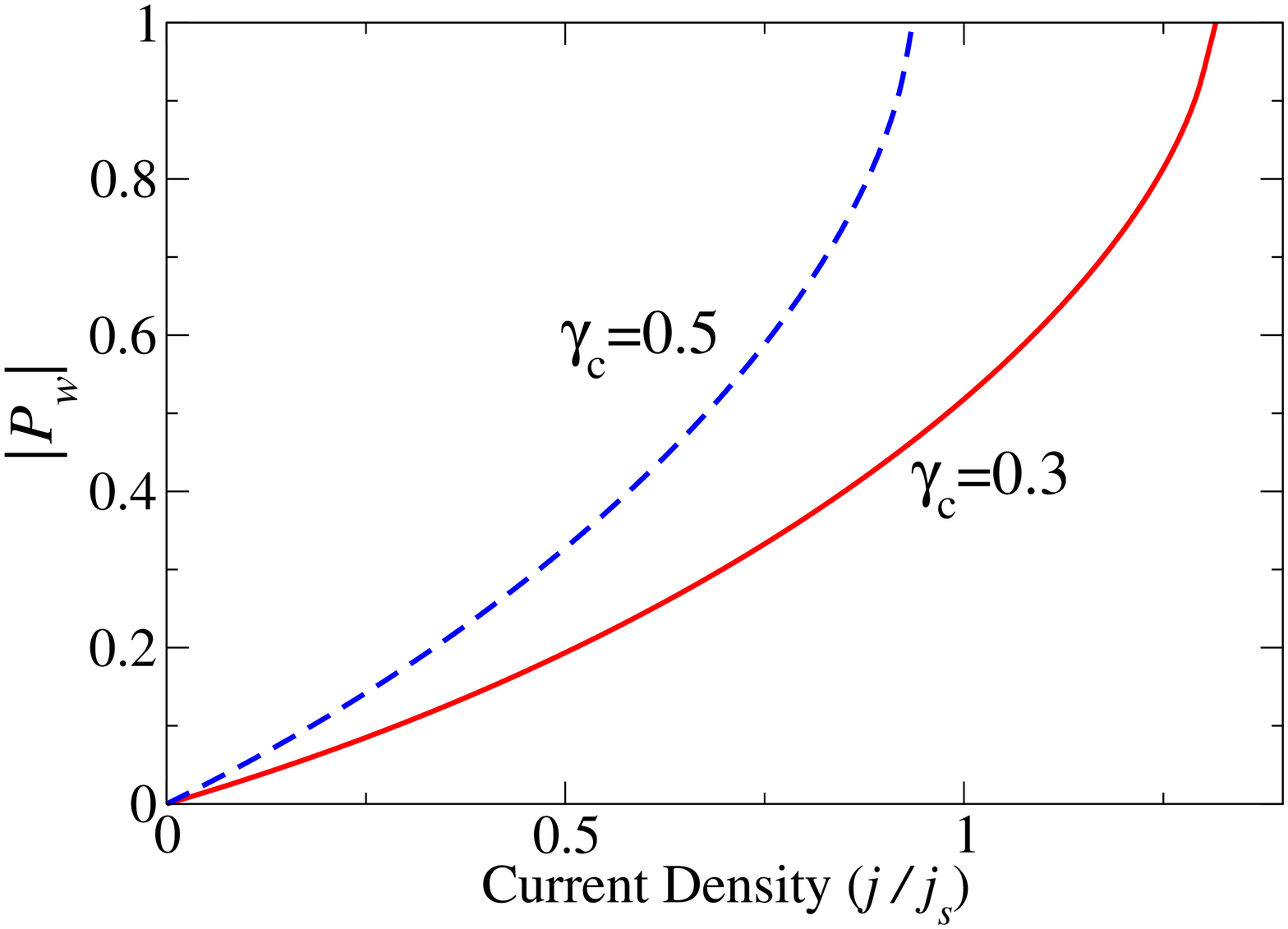}
\caption{(Color online.)
Current dependence of the spin polarization $|P_{w}|$ in $n-S$ at $%
n^{+}/n$ interface}
\label{P0}
\end{figure}

Let us consider the FM-$n^{+}$-$n$ heterostructure (Fig.~\ref{structure})
based on GaAs. The thickness $w$ of the $n^{+}$-layer with electron
concentration $\sim $~10$^{19}$~cm$^{-3}$ is about 10~nm and the electron
concentration $n_{0}$ in $n-S$ region is in the range of 10$^{17}$ - 3$\cdot
$10$^{17}$~cm$^{-3}$. We demonstrate that a 100\%-polarized spin
accumulation layer is formed near $n^{+}-n$ interface $x=w$ when the forward
current density reaches a critical value. The spin-dependent current across
FM-$n^{+}$ interface ($x=0$) can be described by a generalized Landauer
formula \cite{duke1969a}:
\begin{equation}
j_{\sigma }(0)=\frac{e}{4\pi ^{2}h}\int \left[ f(E-\zeta _{\sigma
})-f(E-F_{\sigma })\right] T_{\sigma }(E,\vec{k}_{\parallel },eV)d\vec{k}%
_{\parallel }dE  \label{current}
\end{equation}%
Here $\zeta _{\sigma }$ and $F_{\sigma }$ are the spin-dependent quasi-Fermi
levels in $n^{+}$ and FM layers, respectively. We use the fact that the
splitting of the quasi-Fermi levels in super-heavily doped $n^{+}$ layer is
small compare to the Fermi energy $E_{F}^{+}$ in this region, i.e. $\Delta
\zeta \ll E_{F}^{+}$ and $\Delta \zeta \propto P$. We consider low
temperatures and neglect splitting of the quasi-Fermi levels in the FM
metal. Also we use the local electro-neutrality condition and assume that the
Fermi level of the metal $F=0$. Within this approximation $\zeta _{\sigma
}=eV+\sigma \Delta \zeta /2$, where $\sigma =\pm 1$, and Eq.~(\ref{current})
reads:
\begin{eqnarray}
j_{\sigma }(0) &=&j_{\sigma }^{(0)}(V)+\frac{1}{2}\sigma \Sigma _{\sigma
}(V)\Delta \zeta (0),\text{ where}  \label{current1} \\
j_{\sigma }^{(0)}(V) &=&\frac{e}{4\pi ^{2}h}\int_{\max
\{0,eV-E_{F}^{+}\}}^{eV}T_{\sigma }(E,\vec{k}_{\parallel },eV)d\vec{k}%
_{\parallel }dE  \label{jzero} \\
\Sigma _{\sigma }(V) &=&\frac{e}{4\pi ^{2}h}\int_{\max
\{0,eV-E_{F}^{+}\}}^{eV}T_{\sigma }(eV,\vec{k}_{\parallel },eV)d\vec{k}%
_{\parallel }  \label{Sigma}
\end{eqnarray}%
Taking into account that $\Delta \zeta \propto P<<1$ in the $n^{+}$-layer we
use the standard approximation in which $\Delta \zeta $ in this region
satisfies a linear equation similar to Eq.~(\ref{dims}) with $\tilde{D}=1$
and $j=0$ \cite{aronov1976a,yu2002a,yu2002b,osipov2004a}. By solving this
equation we can express $\gamma(0)$ and $\Delta\zeta(0)$ through
$\gamma(w)$ and $\Delta\zeta(w)$.
The transmission
coefficient of an FM-$n^+$ junction can be represented as $T_{\sigma
}=A_{\sigma }f(\vec{k}_{\parallel },E)$ \cite{osipov2005a}, where $A_{\sigma
}$ is determined by the density of states of electrons with
spin $\sigma $ in
FM and weakly depends on  $E$ and $\vec{k}_{\parallel }$. It
allows us to take $A_\sigma$ out of the integrals in Eqs.~(\ref{jzero}) and
(\ref{Sigma}) and obtain compact expressions for spin extraction coefficient
$\gamma (w)$ and current density $j(V)$:
\begin{eqnarray}
\gamma (w) &=&\frac{j^{(0)}(V)}{j}\left( 1+\frac{\Delta \zeta (w)}{2e}\frac{%
d\ln j^{(0)}(V)}{dV}\right)   \label{gammaw1} \\
j &=&j^{(0)}(V)\left( 1+\gamma _{c}\frac{\Delta \zeta (w)}{2e}\frac{d\ln
j^{(0)}(V)}{dV}\right)   \label{j3}
\end{eqnarray}%
where $\gamma_{c}=\Delta \Sigma /\Sigma$ is the spin selectivity
of the contact \cite{rashba2000a}, $\Sigma =\Sigma _{\uparrow
}+\Sigma _{\downarrow }$, $\Delta \Sigma =\Sigma _{\uparrow }-\Sigma
_{\downarrow }$, and
$j^{(0)}=j_{\uparrow }^{(0)}+j_{\downarrow }^{(0)}$.
Using Eq. (\ref{zi-}) and
matching quasi-Fermi levels at the interface $x=w$ we obtain that in
Eqs.~(\ref{gammaw1}) and (\ref{j3}),
$\Delta \zeta (w)=E_{F}\left[ (1+P_{w})^{2/3}-(1-P_{w})^{2/3}\right]$ ,
where $P_{w}$ is
the spin polarization of the electron density in $n$-S region at $x=w$.
Finally, we use the continuity of $\gamma $ and match
Eq.~(\ref{gammaw1}) with the solution of Eq.~(\ref{gammaP}) in
$n-S$ region. As a
result we obtain spin polarization $P_{w}$, current density $j$,
and spin-extraction
coefficient $\gamma (w)$ as functions of $V$. A typical dependence of $P_{w}$
on $j/j_{s}$ is shown in Fig~\ref{P0}.

\begin{figure}[htbp]
\includegraphics[width=.85\linewidth]{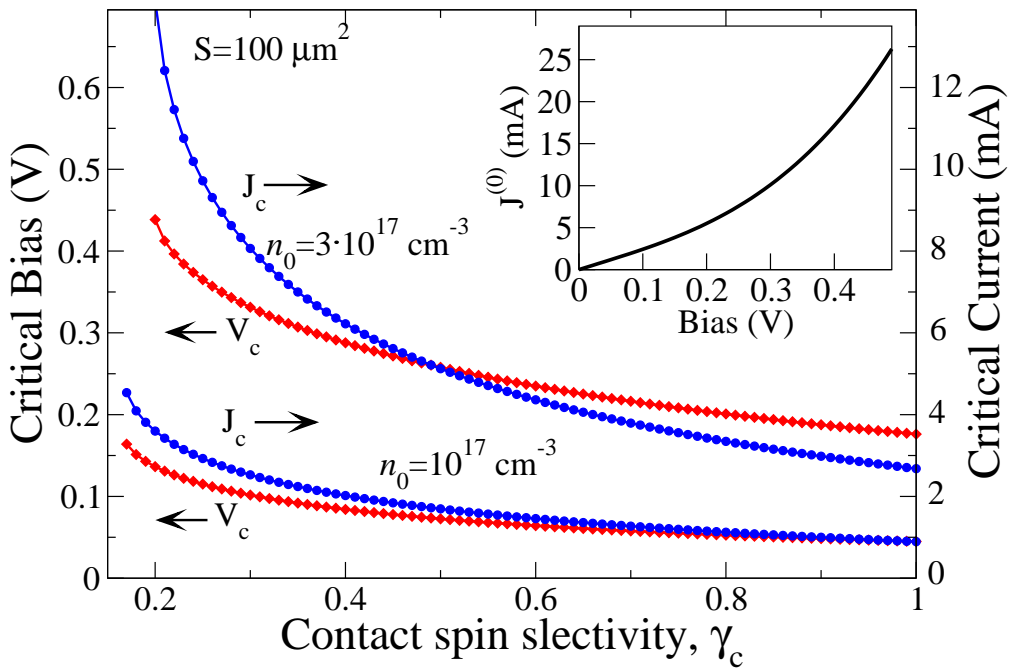} 
\caption{(Color online.)
Critical currents and voltages 
}
\label{threshold}
\end{figure}

The critical current $J_{c}=Sj_{c}$, where $S$ is the
contact area, and voltage $V_{c}$ needed to achieve $%
|P_{w}|=1$ are determined by matching $\gamma $ given by Eqs.~(\ref{gammaw1}%
) and (\ref{gamma7}). The values of $J_{c}$ and $V_{c}$ required to
completely spin polarize electrons of the density $n_{0}$ in $n$-GaAs near $%
n^{+}-n$-interface are shown in Fig.~\ref{threshold}. We used a cubic
approximation for $j^{(0)}(V)$ which is typical for tunnel contacts \cite%
{brinkman1970a} since this approximation is well suited for Fe/GaAs and
Fe/Si tunnel junctions studied experimentally in \cite%
{hanbicki2003a,gareev2003a}. The function $J^{(0)}(V)=Sj^{(0)}(V)$ with
$S=100$~$\mu $m$^{2}$ is shown in the inset to Fig.~\ref%
{threshold}. This function corresponds to a triangular barrier of the height
of 0.63 eV and the effective width of 1.38 nm. We also used $L_{s}=L_{s}^{+}$%
=1~$\mu $m, $\tau _{s}$~=~10$^{-9}$~s, and $w=$~10-30~nm.

We emphasize the crucial role of the $n^{+}$ layer in the proposed FM-$n^{+}$%
-$n$ structure. The presence of $n^{+}$-layer allows us to fabricate a very
thin tunnel barrier which significantly reduces critical
currents and voltages due to its low contact resistance. Moreover, the sharp
concentration drop between the $n^{+}$ and $n$ regions enables a dramatic
change in the spin polarization of the $n$-region while the $n^{+}$-region
is only weakly perturbed. We notice that the transport across the $n^{+}$-$n$
interface is diffusive. At the same time the diffusion coefficient for the
electrons with spin \textquotedblleft up\textquotedblright\ goes to zero.
However, the spatial derivative of the concentration diverges and the
diffusive current remains finite. This effect cannot be realized in 
simple FM-$n-S$ structures where a feedback occurs in the process 
of spin-dependent tunneling
\cite{osipov2004c,bratkovsky2004b,
osipov2005a}.

In conclusion, we emphasize that we have demonstrated the possibility of
achieving 100\% spin polarization in NS via electrical spin extraction,
using FM-$n^{+}$-$n$ structures with moderate spin selectivity.
The highly spin-polarized electrons, according to the results of Refs. \cite%
{kawakami2001a,strand2003a}, can be efficiently utilized to
polarize nuclear spins in semiconductors. They can also be used to
spin polarize electrons on impurity centers or in quantum dots
located near the $n^{+}$-$n$ interface. These effects are
important for spin-based QIP
\cite{awschalom2002a,nielsen2000a,taylor2002a,taylor2003a},
including single electron spin measurements \cite{xiao2004a} and
quantum memory applications \cite{taylor2002a,taylor2003a}. The
considered FM-$n^{+}$-$n$ structures can be used as highly
efficient spin polarizers or spin filters in a majority of the
spin devices proposed to date
\cite{zutic2004a,awschalom2002a,datta1990a,sato2001a,
jiang2003a,osipov2004a,bratkovsky2004a}.
The effect of 100\% spin
polarization can be probed by means of the recently developed spin
trasnport imaging technique \cite{crooker2005a}.

This work was supported by ONR Grant N00014-06-1-0616.
\bibliographystyle{/home/andre/tex/bibtex/apsrev}
\bibliography{/home/andre/tex/bibtex/all}

\end{document}